# Size Dependence of Metal-Insulator Transition in Stoichiometric $Fe_3O_4$ Nanocrystals

Jisoo Lee,[†,‡] Soon Gu Kwon,[†,‡] Je-Geun Park,*[,§,∥] Taeghwan Hyeon*[,†,‡]

[†]Center for Nanoparticle Research, Institute for Basic Science (IBS), Seoul 151-742, Korea

[‡]School of Chemical and Biological Engineering, Seoul National University, Seoul 151-742, Korea

[§]Center for Correlated Electron Systems, Institute for Basic Science (IBS), Seoul 151-747, Korea

[∥]Department of Physics and Astronomy, Seoul National University, Seoul 151-747, Korea

**Table of Contents Graphic**

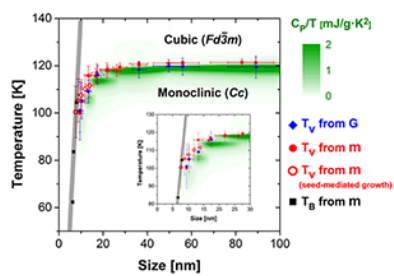

**Abstract**. Magnetite ($Fe_3O_4$) is one of the most actively studied materials with a famous metal-insulator transition (MIT), so-called the Verwey transition at around 123 K. Despite the recent progress in synthesis and characterization of $Fe_3O_4$ nanocrystals (NCs), it is still an open question how the Verwey transition changes on a nanometer scale. We herein report the systematic studies on size dependence of the Verwey transition of stoichiometric $Fe_3O_4$ NCs. We have successfully synthesized stoichiometric and uniform-sized $Fe_3O_4$ NCs with sizes ranging from 5 to 100 nm. These stoichiometric $Fe_3O_4$ NCs show the Verwey transition when they are characterized by conductance, magnetization, cryo-XRD, and heat capacity measurements. The Verwey transition is weakly size-dependent and becomes suppressed in NCs smaller than 20 nm before disappearing completely for less than 6 nm, which is a clear, yet highly interesting indication of a size effect of this well-known phenomena. Our current work will shed new light on this ages-old problem of Verwey transition.

Among various metal oxide materials, $Fe_3O_4$, the oldest known magnetic material, is one of the most actively studied materials with a curie temperature of 858 K. In 1939, Verwey reported that bulk $Fe_3O_4$, which is fairly conductive with a half-metallic character at room temperature, becomes electrically insulating below 123 K, which is now called the Verwey transition temperature ($T_V$).[1] Recently, it was reported that, below the Verwey transition, $Fe^{2+}$ and $Fe^{3+}$ in octahedral sites form a very unusual three-Fe-site 'trimerons' ground state.[2-4] Meanwhile, over the last two decades, iron oxide (magnetite ($Fe_3O_4$) and maghemite ($\gamma$-$Fe_2O_3$)) NCs have been intensively investigated for their various biomedical applications including magnetic resonance imaging (MRI) contrast agents, magnetic biosensors, and heating mediators for magnetic fluid hyperthermia.[5-11] Especially, a recent development in synthesis of uniform and size-controllable NCs[12-14] has enabled the size-dependent physical property characterization and their applications.[15-24] Despite the tremendous progresses in $Fe_3O_4$ NCs,[15, 25-30] it is still an open question how the Verwey transition changes as the particle size gets reduced, in particular on the region of nm scale. The most critical difficulty in this size-dependent characterization of the Verwey transition is the synthesis of uniform and stoichiometric $Fe_3O_4$ NCs[16] because the Verwey transition is reported to be extremely sensitive to oxygen stoichiometry.[31] Most previous studies on the Verwey transition of $Fe_3O_4$ NCs suffer from the difficult problem of oxygen off-stoichiometry, which is an all pervasive issue of every oxide materials including high-temperature superconductors[32] and, more recently, resistive random-access memory (R-RAM)[33] and multiferroics[34] to name only a few.

We synthesized stoichiometric and uniform-sized $Fe_3O_4$ NCs whose diameters ranged from 5 to 100 nm using thermal decomposition of iron acetylacetonate (Fe(acac)$_3$) precursor in the presence of oleic acid surfactant.[12] We could control the particle sizes by two different approaches.

Firstly, we could synthesize Fe$_3$O$_4$ NCs with sizes ranging from 5 nm to 89 nm by varying the precursor-to-surfactant ratios (one-pot thermal-decomposition method) (Figure S2, Supporting Information). Secondly, we employed seed-mediated growth process to synthesize Fe$_3$O$_4$ NCs with sizes of 5 nm ~ 14 nm (Figure S3, Supporting Information).[36] The detailed synthetic processes are provided in the supporting information.

The particle sizes were determined by using transmission electron microscopy (TEM) (Figure 1A-E). TEM images and histograms of NCs show that the particle size distributions are narrow ($\sigma$ < 15%). The high-resolution TEM (HRTEM) images and the selective-area electron diffraction (SAED) patterns (Figure 1F-J) show a highly crystalline structure of the Fe$_3$O$_4$ NCs. The oxidation state of Fe can be qualitatively examined by Fe K-edge X-ray absorption near edge structure (XANES) spectra (Figure 1K).[13] We measured XANES of NCs and compared with iron oxide standards, i.e., maghemite ($\gamma$-Fe$_2$O$_3$) and magnetite (Fe$_3$O$_4$). $\gamma$-Fe$_2$O$_3$ and Fe$_3$O$_4$ have different features near the edge as shown in Figure 1K (more detailed analyses are supplied in Figure S5 in Supporting Information). We confirm that the Fe K-edge peak of NCs are close to Fe$_3$O$_4$ standard.

The HRPD patterns (Figure 1L) show that the peak positions of the Fe$_3$O$_4$ NCs are consistent with those of bulk Fe$_3$O$_4$ (red bars, JCPDS file, No. 19-0629) that carries no sign of other impurities, such as wüstite (FeO).[16] The size and stoichiometry of NCs were estimated by the Rietveld refinement using the Fullprof software as shown in Figure S4 in Supporting Information. The crystal sizes estimated by Rietveld refinement are found to be comparable to the particle sizes measured by TEM (Table S1). Off-stoichiometry parameters ($\delta$), which is defined as Fe$_{3(1-\delta)}$O$_4$, of the NCs are shown in (Table 1). The data show that the NCs are stoichiometric Fe$_3$O$_4$ within error range of 6.5% at maximum. Also, the value of $\delta$ does not show any size dependence which proves that the oxidation state of Fe is not affected by the surface-to-volume ratio.

In order to prepare stoichiometric NCs, we adjusted the electrochemical equilibrium condition of the synthesis by using $CO/CO_2$ gas mixture (Figure S1).[35] In general, there is an equilibrium between $Fe^{2+}$ and $Fe^{3+}$ in iron oxide solid which is often expressed in terms of oxygen fugacity $f_{O_2}$ as follow;[31]

$$4\ Fe_3O_4 + O_2 \rightleftarrows 6\ Fe_2O_3$$

If the reaction condition is reductive so that $f_{O_2}$ is lower than the equilibrium value $f_{O_2}^{\oplus}$, some $Fe^{3+}$ is reduced to $Fe^{2+}$ to restore the equilibrium. In other words, the ratio of $Fe^{2+}$ and $Fe^{3+}$ can be adjusted by the electrochemical potential of oxygen which, in turn, is controlled by the ratio of CO and $CO_2$ in the reaction atmosphere. Because we used $Fe(acac)_3$ as the precursor of iron oxide NCs, there are only $Fe^{3+}$ ions in the solution at the start of the reaction and the exact amount of $Fe^{3+}$ should be reduced to $Fe^{2+}$ to yield stoichiometric $Fe_3O_4$ NCs. In practice, the sharp Verwey transition (Figure 2) is observed in 89 nm- and 36 nm-sized NCs. Because the Verwey transition is extremely sensitive to off-stoichiometry, this result confirms that the synthesis under 4/96 $CO/CO_2$ atmosphere ensures the exact ratio of $Fe^{3+}/Fe^{2+}$ and the formation of stoichiometric $Fe_3O_4$ NCs. It should be noted that the ratio of $Fe^{3+}/Fe^{2+}$ is determined by the electrochemical equilibrium but not by the size of the NCs. Consequently, the NCs with various sizes synthesized under 4/96 $CO/CO_2$ atmosphere should have the same $Fe^{3+}/Fe^{2+}$ ratio and stoichiometry as 89 nm and 36 nm-sized NCs.

It is widely known that the Verwey transition of $Fe_3O_4$ is accompanied by various anomalies in the physical properties at $T_V$,[1, 2, 37] including magnetic moment (m), conductance (G), structural (space group) change, and heat capacity ($C_P$). First, we examined the structural change of NCs at the Verwey transition by using a cryo-XRD (Figure S6, Supporting Information). For bulk $Fe_3O_4$, there is a drastic change in the crystal symmetry from monoclinic (Cc) below $T_V$ to inverse spinel cubic (Fd$\bar{3}$m) above $T_V$.[2] For example, the (440) peak in the cubic phase is split into two peaks below $T_V$ with a reduction in the peak intensity in the low-temperature monoclinic phase[38, 39] as shown in the data for bulk $Fe_3O_4$, which is absent in $\gamma$-$Fe_2O_3$ (Figure S6, Supporting Information). This change is still visible in NCs, for instance, in 22 nm-sized NCs, and becomes considerably weaker as the particle size decreases.

Furthermore, an abrupt change in the conductance is the clear evidence of the Verwey transition, i.e. a metal-insulator transition (MIT). In the picture of charge ordering originally put forward by Verwey himself and further refined by more recent researches,[2, 3] $Fe^{2+}$ and $Fe^{3+}$ are ordered into a unique linear three-Fe-site unit below $T_V$. These localized and charge-ordered electrons organize themselves making $Fe_3O_4$ electrically insulating below $T_V$. Above $T_V$, electrons are no more localized and can hop between $Fe^{2+}$ and $Fe^{3+}$ sites. In order to examine the size dependence of the MIT, we have measured the conductance of NCs to find that the sign of MIT gets progressively weaker as the particle size is reduced. In Figure 2A, the conductance (G) is plotted as $d(\ln G)/d(1000/T)^{1/4}$ vs $(1000/T)^{1/4}$: the minimum is clearly visible even in the data of 10 nm although the anomaly is much more suppressed and a lot broader than the minimum of bulk $Fe_3O_4$.

When $Fe_3O_4$ undergoes the MIT, there is also an anomaly in the magnetization data with a change in the magnetic easy axis: $Fe_3O_4$ is a spinel type ferrimagnetic material with $Fe^{3+}$ at the tetrahedral site (A) and $Fe^{2+}/Fe^{3+}$ at the octahedral site (B) having a strong antiferromagnetic interaction. For instance, bulk $Fe_3O_4$ shows a drastic drop in the magnetization data at 123 K shown in Figure 2B, where we observed the MIT in the conductance measurement and the structure change, consistent with the reported results,[37] which renders it another useful test of the Verwey transition in NCs. As clearly seen in our data, this anomaly in the magnetization moves only slightly toward lower temperatures and gets broader as the particle size decreases. Intriguingly, the Verwey transition disappears below 6 nm, where blocking temperature ($T_B$) becomes lower than $T_V$ (Figure 2B, Figure S7 in Supporting Information).

Although the anomalies we observed in the cryo-XRD, conductance, and magnetization data are conspicuous, it is the heat capacity data that can confirm these anomalies as a truly thermodynamic transition as in the bulk $Fe_3O_4$. For example, it is possible that $T_V$ is masked by rapid decrease in the magnetic moment in the temperature range below blocking temperature ($T_B$). Thus, it is reassuring the heat capacity data measured on NCs shows a lambda-like anomaly just like that of bulk $Fe_3O_4$ (Figure 2C). From this, we can deduce three statements. First and foremost, all the anomalies we observed in the NCs are a true Verwey transition on a nanometer scale, whose peak gets progressively broader while it moves toward lower-temperature. Second, the total area of the peak, corresponding to the total entropy of the Verwey transition for a given size of samples, shows a reduction as the particle size gets smaller although the transition temperature itself has barely changed, which indicates that lesser entropy is released for the Verwey transition of smaller NCs. Third, the peak of the Verwey transition is completely absent in the 6 nm-sized NCs (Figure S8, Supporting Information), indicating that the Verwey transition disappears in $Fe_3O_4$ NCs smaller than 6 nm.

Based on thermodynamic consideration, we stated above that the size of the NCs does not affect the ratio of $Fe^{3+}$ and $Fe^{2+}$ in equilibrium with the $CO/CO_2$ atmosphere. In order to support this statement and stoichiometry issue experimentally, we designed synthesis experiments using seed-mediated growth method. In these experiments, we used 5 nm NCs as the seeds to prepare 6, 8, 10, and 14 nm NCs vis sequential seed-mediated growth method (Figure 3A).[36, 40] In this way, we obtained the larger NCs with their core having the stoichiometry of the smaller seed NCs. If the size of the NCs affects the $Fe^{3+}/Fe^{2+}$ ratio, then the Verwey transition behaviors of the NCs prepared by seed-mediated growth method and by the one-pot synthesis method (heating $Fe(acac)_3$ solution without the seeds) should be different. As shown in Figure 3B, zero-field-cooled (ZFC) magnetization data reveal that there are clear stepwise magnetic moment changes near 120 K in the 10 nm and 14 nm samples, which coincide almost perfectly with the data from the NCs with the same sizes but prepared by the one-pot thermal-decomposition method. This observation confirms that the size of the NCs has nearly no effect on the stoichiometry of $Fe_3O_4$ NCs and the size-dependent behavior of the Verwey transition is reproducible with good reliability from the samples prepared via two different synthetic methods.

Thus, by combining our data taken on all the samples prepared using the two methods in a wide-ranging size, we can confirm that the Verwey transition is size-independent down to 20 nm with a weak suppression of the transition temperature from 20 to 8 nm before suddenly disappearing below 6 nm (Figure 4). Interestingly, the size vs. the Verwey transition temperature plot is well fitted with an exponential equation, $T_V = a \cdot \exp(-D/b) + T_o$ where $a$, $b$, and $T_o$ are fit parameters (Figure S10, Supporting Information). The adjusted $R^2$ value of 0.937 from this fit is statistically meaningful and it is strongly suggested that there is quantitative relationship between the size of the NCs and $T_V$. There is a possibility that broken symmetry of the crystal structure and the presence of low-coordinated atoms at the surface of the NCs can affect the observed size-dependent behavior of the Verwey transition, which becomes more pronounced for the smaller

NCs. Also this result excludes the possibility of contribution from surface oxidation and disorder to the Verwey transition. If the stoichiometry of the NCs is changed by any effect from the surface, it is reasonable to expect that off-stoichiometry is proportional to the surface-to-volume ratio. That is, the off-stoichiometry parameter $\delta$ is proportional to $1/D$ where $D$ is the size of the NC. Then, because the Verwey transition temperature $T_V$ is linearly dependent on $\delta$,[31] we should have $T_V \propto 1/D$, which is rejected by the exponential fit result in Figure S10 in Supporting Information.

This drastic size effect of the Verwey transition has never been reported before and reveals an unexpected yet intimate inner secret of this decades-old conundrum. As the history of seven decades-long researches shows, the origin of the Verwey transition is an extremely difficult problem and inevitably our ideas for this just uncovered experimental observation is bound to be speculative. Nevertheless, let us offer our thoughts that might well be found later useful for a full-fledge theoretical investigation. First, a commonly accepted explanation is that the spin degree of freedom is irrelevant for the Verwey transition unlike the other two degrees of freedom like charge and orbital. In this regards, it is interesting to note that the Verwey transition appears to suffer a drastic disappearance when the line of the transition temperature in Figure 4 hits a line of blocking temperature ($T_B$) (thick gray line in Figure 4), below which the giant magnetic moment of the single magnetic domain of the whole particle experiences a freezing of a thermally assisted over-the-barrier transition. Therefore, we propose with caution that after all the spin degree of freedom may not be that irrelevant for the Verwey transition. Second, we have demonstrated that the Verwey transition shows the remarkable size effect which might as well prompt one to think of carrying out full unbiased first-principles calculations on $Fe_3O_4$ nanoparticles with size just below and above the critical value of 6 nm. This new, nonetheless very challenging theoretical study will throw then the definite and ultimate light on our understanding of the seven decades long riddle.

In this work, we successfully synthesized uniform and, more importantly, stoichiometric $Fe_3O_4$ NCs down to 5 nm, and undertook in-depth studies of the size-dependence of the Verwey transition, which is of particular significance for the fundamental understanding of the issue. These comprehensive experimental investigations combined several microscopic and bulk characterization tools: X-ray absorption spectroscopy, conductance, X-ray diffraction, magnetization and heat capacity. Based on these data, we conclude that there is a critical size for the Verwey transition at around 6 nm. The observation of the drastic size effect of the Verwey transition reported here throws a fresh glimpse into the decades-long conundrum and offers a new yet challenging window of opportunities.

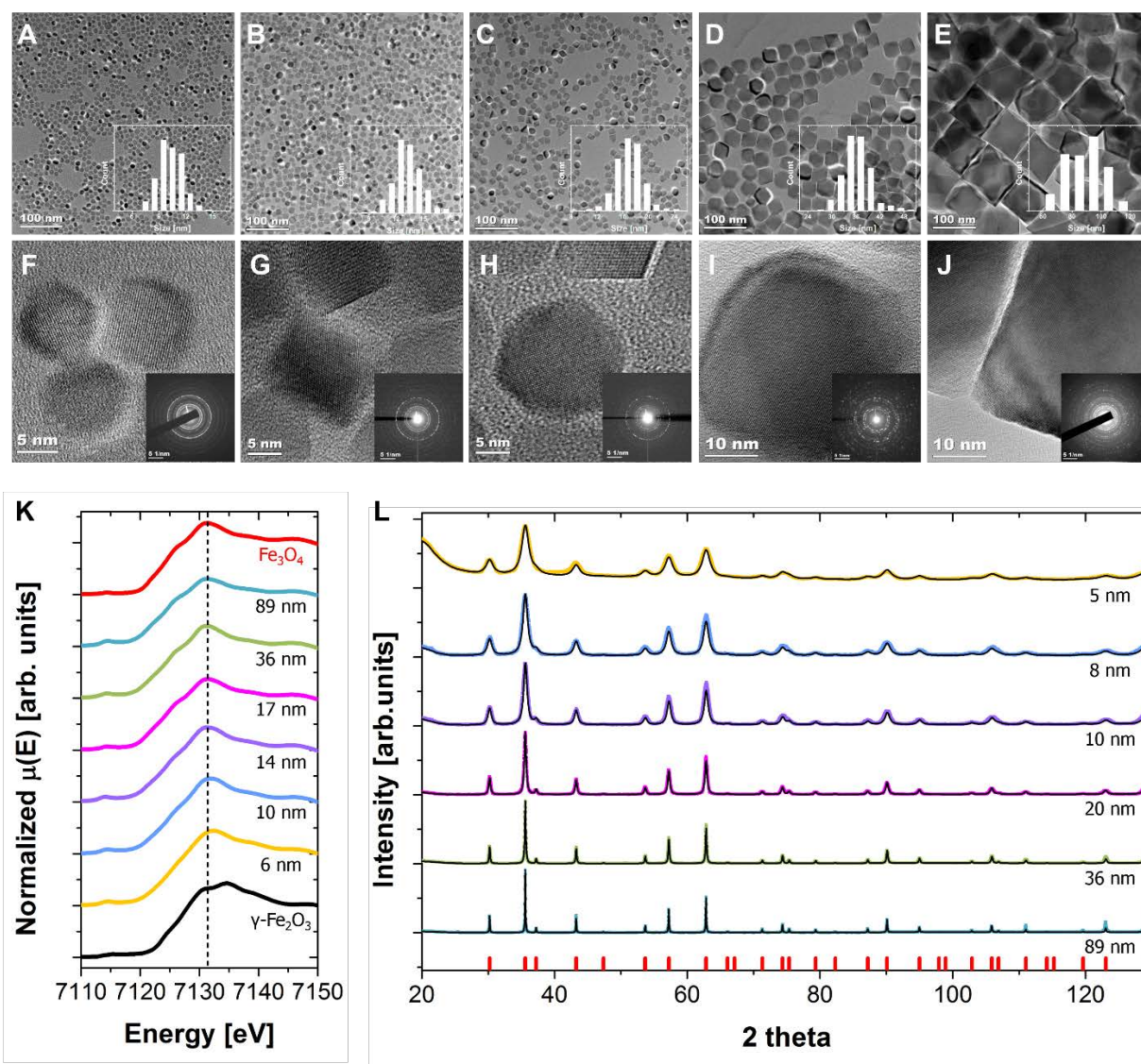

**Figure 1.** Characterization of $Fe_3O_4$ nanocrystals. TEM images (A-E) and HRTEM images (F-J) of $Fe_3O_4$ NCs. Particle size distribution histograms are shown in the inset of (A-E) with SAED patterns in the inset of (F-J): (A, F) 10 nm, (B, G) 14 nm, (C, H), 17 nm, (D, I) 36 nm, (E, J) 89 nm. The histograms show narrow particle size distributions while the HRTEM images with SAED patterns attest to the high crystallinity. (K) Fe K-edge XANES spectra of $Fe_3O_4$ NCs and bulk $\gamma$-$Fe_2O_3$ and $Fe_3O_4$ standard samples. The dotted line indicated the white line peak position of bulk $Fe_3O_4$. (L) HRPD patterns of $Fe_3O_4$ NCs overlapped with profile functions of Rietveld refinement (black).

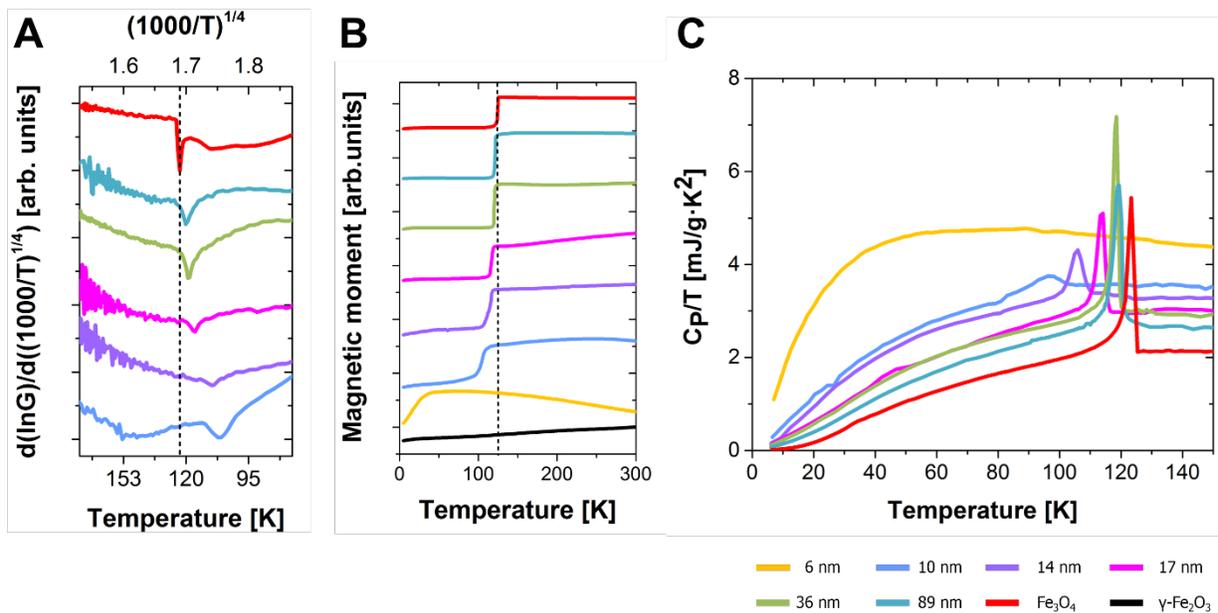

**Figure 2.** Size dependent characterization of the Verwey transition. (A) Temperature dependence of first derivative of conductance (G). The sharp peaks correspond to the Verwey transition, still visible down to 10 nm although it shifts toward lower temperature with decreasing particle size. (B) Zero-field-cooled (ZFC) magnetization data measured at magnetic field of 10 mT for various sized $Fe_3O_4$ NCs. The sharp drop in the magnetization near 120 K indicates the Verwey transition. The vertical dotted line indicates $T_V$ of bulk $Fe_3O_4$. (C) The total heat capacity divided by temperature. As the particle size decreases, the sharp peak of the Verwey transition moves toward lower temperature and gets considerably weaker.

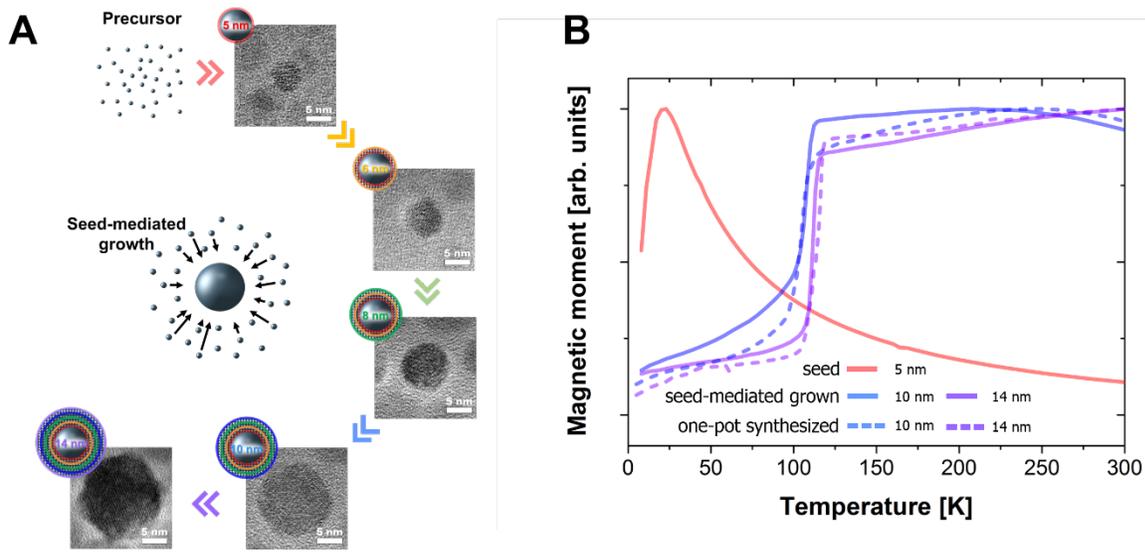

**Figure 3.** The evolution of the Verwey transition in $Fe_3O_4$ NCs with sizes from 5 nm to 14 nm. (A) $Fe_3O_4$ NCs with sizes of 6, 8, 10, and 14 nm were synthesized using 5 nm-sized NCs as seeds. (B) Zero-field-cooled (ZFC) magnetization data is measured under field of 10 mT. The step-wise magnetic moment change near 120 K indicates the Verwey transition. Note that the $T_V$ of seed-mediated grown NCs (solid line) and one-pot synthesized NCs (dotted line) are nearly identical.

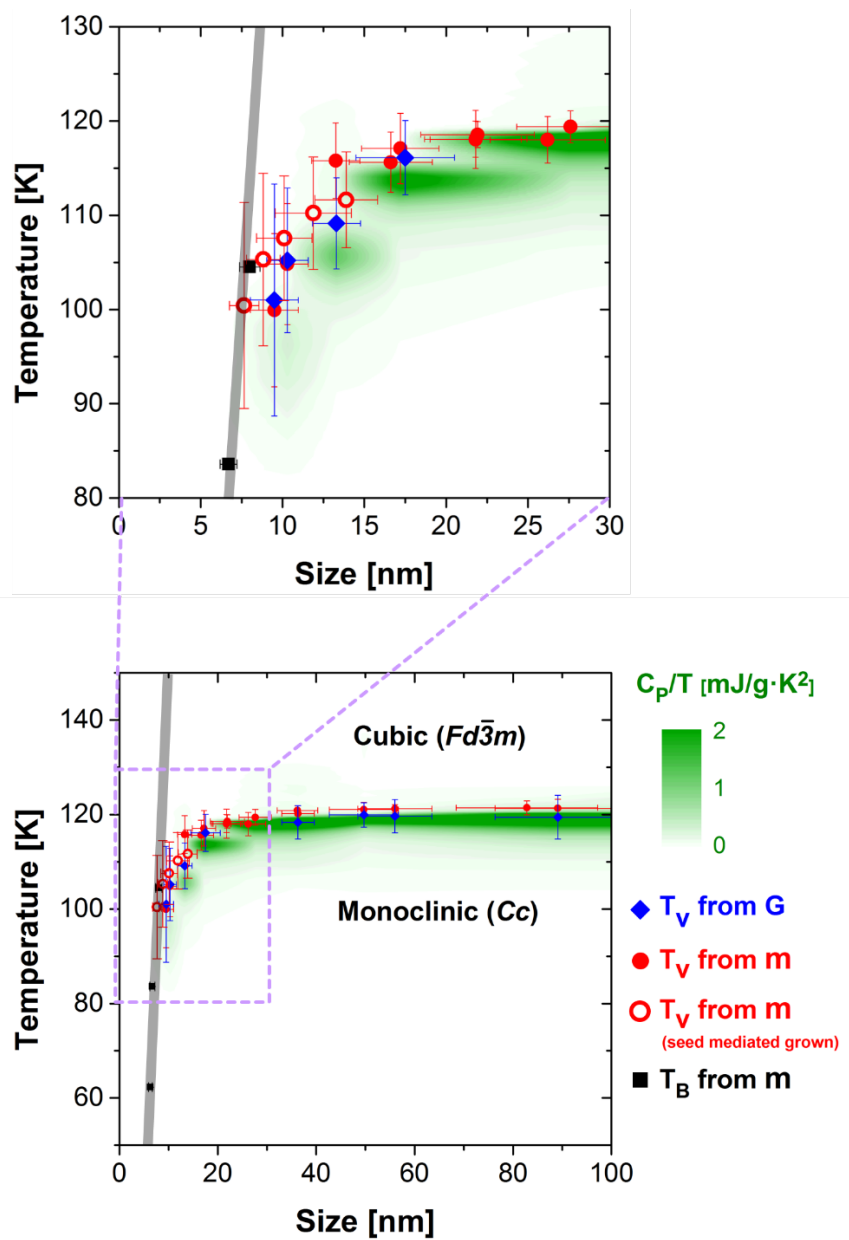

**Figure 4.** Size dependence of $T_V$ for $Fe_3O_4$ NCs. The contour plot represents the heat capacity data after removing the contribution of surfactant (See Figure S8 in Supporting Information). The symbols mark the Verwey transition temperature ($T_V$) determined from three different types of measurements: heat capacity (green, $C_P/T$), conductance (blue, G), and magnetic moment (red, m). The size dependence of blocking temperature ($T_B$) is also plotted from the same magnetization measurement (black). The gray line is a guide to the eye for the size dependence of $T_B$.

**Table 1.** Off-stoichiometry parpameter δ of the NCs and corresponding reliable factors, $R_B$, $R_F$, and $\chi^2$ derived by Rietveld refinement of HRPD data.

| Size [nm] | $R_B$ [%] | $R_F$ [%] | $\chi^2$ | $Fe_{3(1-\delta)}O_4$ |
|:---:|:---:|:---:|:---:|:---:|
| 5 | 11.0 | 8.36 | 16.4 | δ ~ 0.018 |
| 8 | 6.54 | 3.51 | 5.60 | δ ~ 0.058 |
| 10 | 8.46 | 6.17 | 7.14 | δ ~ 0.065 |
| 20 | 6.12 | 4.23 | 6.82 | δ ~ 0.030 |
| 36 | 9.50 | 7.14 | 8.08 | δ ~ 0.016 |
| 89 | 19.6 | 10.6 | 3.90 | δ ~ 0.004 |

## ASSOCIATED CONTENT

**Supporting Information**

Experimental details, temperature dependence of magnetization curves, TEM images and size distribution histograms, X-ray HRPD patterns, XANES spectra, thermal evolution of (440) plane XRD peak, zero-field-cooled and field-cooled magnetization data, heat capacity curves, refined result of HRPD. This material is available free of charge via the Internet at http://pubs.acs.org.

**Corresponding Authors**

*E-mail: jgpark10@snu.ac.kr

*E-mail: thyeon@snu.ac.kr

**Notes**

The authors declare no competing financial interest.

**Acknowledgements**

This work was supported by the Institute for Basic Science (IBS) in Korea (IBS-R006-D1 and IBS-R009-G1). We thank Y. M. Song and I. S. Lee of the National Center for Inter-University Research Facilities (NCIRF) of Seoul National University for assistance with the PPMS and TEM experiments. We thank Prof. Jae-Hoon Park of Pohang University of Science and Technology (POSTECH) for helpful discussions. We also thank S. I. Choi for technical assistance to use a home-built resistance set-up. Experiments at PLS were supported in part by MSIP and POSTECH.

**References**

1. Verwey, E. J. *Nature* **1939,** *144*, 327-328.
2. Senn, M. S.; Wright, J. P.; Attfield, J. P. *Nature* **2012,** *481*, 173-176.
3. de Jong, S.; Kukreja, R.; Trabant, C.; Pontius, N.; Chang, C. F.; Kachel, T.; Beye, M.; Sorgenfrei, F.; Back, C. H.; Bräuer, B.; Schlotter, W. F.; Turner, J. J.; Krupin, O.; Doehler, M.; Zhu, D.; Hossain, M. A.; Scherz, A. O.; Fausti, D.; Novelli, F.; Esposito, M.; Lee, W. S.; Chuang, Y. D.; Lu, D. H.; Moore, R. G.; Yi, M.; Trigo, M.; Kirchmann, P.; Pathey, L.; Golden, M. S.; Buchholz, M.; Metcalf, P.; Parmigiani, F.; Wurth, W.; Föhlisch, A.; Schüßler-Langeheine, C.; Dürr, H. A. *Nature Mater.* **2013,** *12*, 882-886.
4. Senn, M. S.; Loa, I.; Wright, J. P.; Attfield, J. P. *Phys. Rev. B* **2012,** *85*, 125119.
5. Laurent, S.; Forge, D.; Port, M.; Roch, A.; Robic, C.; Vander Elst, L.; Muller, R. N. *Chem. Rev.* **2008,** *108*, 2064-2110.
6. Xie, J.; Liu, G.; Eden, H. S.; Ai, H.; Chen, X. *Acc. Chem. Res.* **2011,** *44*, 883-892.
7. Stanley, S. A.; Gagner, J. E.; Damanpour, S.; Yoshida, M.; Dordick, J. S.; Friedman, J. M. *Science* **2012,** *336*, 604-608.
8. Lee, N.; Hyeon, T. *Chem. Soc. Rev.* **2012,** *41*, 2575-2589.
9. Lee, N.; Choi, Y.; Lee, Y.; Park, M.; Moon, W. K.; Choi, S. H.; Hyeon, T. *Nano Lett.* **2012,** *12*, 3127-3131.
10. Jun, Y.-w.; Seo, J.-w.; Cheon, J. *Acc. Chem. Res.* **2008,** *41*, 179-189.
11. Jun, Y.-w.; Lee, J.-H.; Cheon, J. *Angew. Chem. Int. Ed.* **2008,** *47*, 5122-5135.
12. Sun, S.; Zeng, H. *J. Am. Chem. Soc.* **2002,** *124*, 8204-8205.
13. Park, J.; An, K.; Hwang, Y.; Park, J.-G.; Noh, H.-J.; Kim, J.-Y.; Park, J.-H.; Hwang, N.-M.; Hyeon, T. *Nature Mater.* **2004,** *3*, 891-895.
14. Ho, D.; Sun, X.; Sun, S. *Acc. Chem. Res.* **2011,** *44*, 875-882.
15. Zeng, H.; Black, C. T.; Sandstrom, R. L.; Rice, P. M.; Murray, C. B.; Sun, S. *Phys. Rev. B* **2006,** *73*, 020402.
16. Redl, F. X.; Black, C. T.; Papaefthymiou, G. C.; Sandstrom, R. L.; Yin, M.; Zeng, H.; Murray, C. B.; O'Brien, S. P. *J. Am. Chem. Soc.* **2004,** *126*, 14583-14599.
17. Couchman, P. R.; Jesser, W. A. *Nature* **1977,** *269*, 481-483.
18. Goldstein, A. N.; Echer, C. M.; Alivisatos, A. P. *Science* **1992,** *256*, 1425-1427.
19. Jacobs, K.; Zaziski, D.; Scher, E. C.; Herhold, A. B.; Alivisatos, A. P. *Science* **2001,** *293*, 1803-1806.
20. Chen, C.-C.; Herhold, A. B.; Johnson, C. S.; Alivisatos, A. P. *Science* **1997,** *276*, 398-401.
21. Lang, X. Y.; Zheng, W. T.; Jiang, Q. *Phys. Rev. B* **2006,** *73*, 224444.
22. Rao, C. N. R.; Kulkarni, G. U.; Thomas, P. J.; Edwards, P. P. *Chem. Eur. J.* **2002,** *8*, 28-35.
23. Polking, M. J.; Han, M.-G.; Yourdkhani, A.; Petkov, V.; Kisielowski, C. F.; Volkov, V. V.; Zhu, Y.; Caruntu, G.; Alivisatos, A. P.; Ramesh, R. *Nature Mater.* **2012,** *11*, 700-709.
24. Banin, U.; Cao, Y.; Katz, D.; Millo, O. *Nature* **1999,** *400*, 542-544.
25. Lu, A.-H.; Salabas, E. L.; Schüth, F. *Angew. Chem. Int. Ed.* **2007,** *46*, 1222-1244.
26. Zhang, D.; Liu, Z.; Han, S.; Li, C.; Lei, B.; Stewart, M. P.; Tour, J. M.; Zhou, C. *Nano Lett.* **2004,** *4*, 2151-2155.
27. Lee, S.; Fursina, A.; Mayo, J. T.; Yavuz, C. T.; Colvin, V. L.; Sumesh Sofin, R. G.; Shvets,

I. V.; Natelson, D. *Nature Mater.* **2008,** *7*, 130-133.
28. Disch, S.; Wetterskog, E.; Hermann, R. l. P.; Salazar-Alvarez, G.; Busch, P.; Brückel, T.; Bergström, L.; Kamali, S. *Nano Lett.* **2011,** *11*, 1651-1656.
29. Kim, T. H.; Jang, E. Y.; Lee, N. J.; Choi, D. J.; Lee, K.-J.; Jang, J.-t.; Choi, J.-s.; Moon, S. H.; Cheon, J. *Nano Lett.* **2010,** *10*, 2734-2734.
30. Noh, S.-h.; Na, W.; Jang, J.-t.; Lee, J.-H.; Lee, E. J.; Moon, S. H.; Lim, Y.; Shin, J.-S.; Cheon, J. *Nano Lett.* **2012,** *12*, 3716-3721.
31. Shepherd, J. P.; Koenitzer, J. W.; Aragón, R.; Spałek, J.; Honig, J. M. *Phys. Rev. B* **1991,** *43*, 8461-8471.
32. Bednorz, J. G.; Müller, K. A. *Z. Physik B - Condensed Matter* **1986,** *64*, 189-193.
33. Waser, R.; Aono, M. *Nature Mater.* **2007,** *6*, 833-840.
34. Lee, S.; Pirogov, A.; Kang, M.; Jang, K.-H.; Yonemura, M.; Kamiyama, T.; Cheong, S. W.; Gozzo, F.; Shin, N.; Kimura, H.; Noda, Y.; Park, J. G. *Nature* **2008,** *451*, 805-808.
35. Kang, Y.; Ye, X.; Murray, C. B. *Angew. Chem. Int. Ed.* **2010,** *49*, 6156-6159.
36. Park, J.; Lee, E.; Hwang, N.-M.; Kang, M.; Kim, S. C.; Hwang, Y.; Park, J.-G.; Noh, H.-J.; Kim, J.-Y.; Park, J.-H.; Hyeon, T. *Angew. Chem. Int. Ed.* **2005,** *44*, 2872-2877.
37. Friedrich, W. *J. Phys.: Condens. Matter* **2002,** *14*, R285.
38. Wright, J. P.; Attfield, J. P.; Radaelli, P. G. *Phys. Rev. B* **2002,** *66*, 214422.
39. Blasco, J.; García, J.; Subías, G. *Phys. Rev. B* **2011,** *83*, 104105.
40. Sun, S.; Zeng, H.; Robinson, D. B.; Raoux, S.; Rice, P. M.; Wang, S. X.; Li, G. *J. Am. Chem. Soc.* **2004,** *126*, 273-279.